\documentclass{article}
\usepackage{graphicx} 

\title{The Reference System Loophole}
 \author{Emily Adlam \thanks{Philosophy Department and Institute for Quantum Studies, Chapman University, Orange, CA92866, USA \texttt{eadlam90@gmail.com} }}
\date{\today}
\usepackage{amsmath}
\usepackage{natbib} 
 \usepackage{gensymb} 

\begin{document}

\maketitle

\begin{abstract}
 I argue that the Extended Wigner's Friend theorems have a `loophole' that has previously not been appreciated, coming from  the fact that we typically need a reference system to give meaning to a measurement outcome. I show how this loophole can be used in the context of the Local Friendliness theorem to preserve the predictions of quantum mechanics without denying Local Friendliness, and I respond to some objections. I explain how  the loophole arises naturally from consideration of relevant symmetry transformations, and then I appeal to the partial/complete observables formalism to argue that the same loophole will work for any type of Wigner's Friend experiment, even if it is performed with a variable that does not obviously look like it requires relativization to a reference system. Finally I compare the loophole to other possible responses to the Local Friendliness theorem, and I show that the same loophole will work for other Extended Wigner's Friend theorems too.

\end{abstract}

\section{Introduction}

In recent years there have  been exciting developments in the study of Extended Wigner's Frame experiments aiming to probe the nature of observation in quantum mechanics. In this article I will argue that the Extended Wigner's Friend theorems have a `loophole' that has previously not been appreciated, coming from  the fact that we typically need a reference system to give meaning to a measurement outcome. I suggest that this loophole has interesting connections to ongoing research programmes about symmetries and reference frames in quantum physics.

What I will call the `reference system loophole' arises because standard treatments of the EWF paradoxes assume that the variable describing what Alice sees in her laboratory is the same as the variable expressed relative to Bob's lab that we will ultimately compare to the  predictions of quantum mechanics, but in fact the two can be identified in this way only if Alice's reference system always stands in the same well-defined relation  to Bob's reference system. Distinguishing the  two variables thus provides a possible way to preserve the  predictions of unitary quantum mechanics without denying the other assumptions used to prove the theorems.

This is important first and foremost because these theorems are   \emph{no-go theorems}, which means that their purpose is to demonstrate a contradiction between  some set of plausible assumptions. For example, the Local Friendliness theorem of \cite{Bong_2020} is usually glossed as proving that if the predictions of quantum mechanics are universally correct then locality, no-superdeterminism or aboluteness of observed events must be violated. Thus in considering how to respond to these theorems it is essential to  uncover any additional assumptions used in the proof, in order to properly understand the full range of possible responses. The reference system assumption is not mentioned in the original Bong et al paper or in any of the subsequent literature, so there is clearly a need for it to be stated more explicitly. Moreover, even if one finds the idea of violating the reference system assumption  implausible or unattractive, I will argue that it is at least more appealing  than violating absoluteness of observed events, which is the approach that has  dominated much of the literature on the Local Friendliness theorem. 

To put the current paper in context, this project should be understood in the same spirit as earlier work exploring the logical structure of Bell's theorem and uncovering various implicit assumptions of the theorem, such as proposals for retrocausal or superdeterministic responses \citep{deBeauregard1953-DEBMQ,hossenfelder2020superdeterminism,RevModPhys.92.021002}. I emphasize that, just like comparable proposals pertaining to Bell's theorem, the approach described here requires modifications to standard quantum mechanics. In particular, although the proposal arises from taking seriously the role of reference systems and draws on some ideas from the quantum reference frame (QRF) formalism, the proposal is not intended to be implemented  within the existing QRF approach - in section \ref{superposition} we will see in more detail why the QRF formalism cannot be used for this purpose.

  I will begin in section \ref{Wigner} by first introducing the Local Friendliness theorem of \cite{Bong_2020}. Then in section \ref{loophole} I will explain how the reference system loophole works for the specific case where the Bong et al experiment is performed using  measurements of the orientation of spin, and I will respond to a few obvious objections. In section \ref{symmetry} I will give a more detailed discussion of the way in which consideration of   symmetries makes the reference system loophole   natural, and in  section \ref{otherobservables} I will discuss the prospects for the approach to be extended to versions of the experiment involving variables other than spin. Finally in section \ref{why} I will compare the reference system loophole to some alternative  responses to the Local Friendliness theorem, and I will briefly explain how this loophole can also work to resolve other kinds of EWF experiments. 
 
\section{Wigner's Friend \label{Wigner}}
 
The standard Wigner's Friend scenario \citep{Wigner} involves an observer, Alice, equipped with a qubit $S$  in the state $x | \downarrow \rangle + y |\uparrow \rangle$, with both $x$ and $y$ non-zero. Suppose Alice performs a spin measurement with possible outcomes `up' ($\uparrow$) and `down' ($\downarrow$) associated with the operators $\{ |\uparrow \rangle \langle \uparrow |, |\downarrow \rangle \langle \downarrow | \}$.  Alice presumably  obtains some outcome to her measurement,  and   quantum mechanics as standardly applied tells us that the system is now in the state corresponding to that outcome:   if Alice sees the outcome `up' then the state of system $S$ is now $| \uparrow \rangle$, and if Alice sees the outcome `down' then the  state of system $S$ is now $|\downarrow \rangle$. 

However, suppose that Bob, observing from the outside, is able to maintain Alice's laboratory system in a coherent state, and he describes the whole interaction using unitary quantum mechanics. He will conclude that  $S$ and Alice are now in the following entangled superposition state: 

\[ \psi_{SA} = x | \uparrow \rangle_S | see \ up \rangle_A + y |\downarrow \rangle_S | see \ down \rangle_A \]

Here, $|see \ up \rangle_A$ and $| see \ down \rangle_A$ are states of Alice corresponding to `seeing the up outcome' and `seeing the down outcome' respectively. 

Something strange seems to be happening:  presumably Alice has seen a single outcome, but the state assigned by Bob appears to represent Alice as somehow experiencing a combination of two different outcomes. Moreover, if we think quantum mechanics is universally correct in its predictions we cannot say that the state relative to Bob just represents his lack of knowledge of the true outcome, for the universality of quantum mechanics implies that if Bob measures Alice and $S$ together he will  see outcomes corresponding to the entangled state $\psi_{SA}$ rather than  finding $S$ in one of the pure states $|\downarrow\rangle_S$ or $| \uparrow \rangle_S$, so in some robust sense  $\psi_{SA}$  is really the correct state relative to him.

The worry here can be sharpened by considering an `Extended Wigner's Friend' (EWF) scenario which combines two Wigner's Friend experiments with a Bell experiment. There are various theorems of this kind, but in this paper I will largely focus on the Local Friendliness theorem due to \cite{Bong_2020}; other EWF theorems are discussed in section \ref{EWF}. 

In the Bong et al experiment we prepare an entangled state on two systems which I will call $S_A$ and $ S_C$. We then give $S_A$ to Alice and have her perform a fixed measurement on it, obtaining outcome $A$, and we give $S_C$ to Chidi and have him perform a fixed measurement on it, obtaining outcome $C$. Meanwhile Bob keeps Alice's laboratory in a coherent state and subsequently performs on it a `supermeasurement' which does not commute with Alice's original measurement, obtaining an outcome $B$, while Divya does the same with Chidi, obtaining an outcome $D$. Note that in order for Bob to be able to perform the supermeasurement it is essential that he prevents Alice from undergoing any decoherence relative to his laboratory, which means that Alice must be kept very well isolated from the external world and no information about her measurement outcome can be allowed to escape her laboratory until Bob performs his measurement.

\cite{Bong_2020} use this setup to prove the `Local Friendliness theorem,' showing that if  we assume the existence of a well-defined probability distribution over these four outcomes which obeys a certain locality condition and which does not involve any superdeterminism or retrocausality (a conjunction of assumptions sometimes referred to as `Local Friendliness'),  this distribution must obey a set of  inequalities known as the Local Friendliness (LF) inequalites. Then they show that for certain choices  of entangled state and measurements, there is no probability distribution over the four measurement outcomes $A, B, C, D$ which satisfies the LF inequalities while reproducing the pairwise predictions of quantum mechanics for  $\{ (B,D), (B, C), $ $(A, D), (A, C) \}$.  Specifically, to demonstrate this they employ  two-qubit photon polarization states expressed in terms of horizontal and vertical polarization states all stated relative to the same axis. So Local Friendliness appears to be inconsistent with the predictions of quantum mechanics, if we assume that quantum mechanics applies universally. 

Now, it should be noted that Bob necessarily erases Alice's outcome $A$ when performing the measurement to find out the value of $B$, and likewise Divya erases $C$ when performing a measurement to find out the value of $D$, so initially it looks as though the only pair of measurements that anyone can actually compare is $(B, D)$. Thus one might try to avoid the paradox by proposing that the predictions of quantum mechanics can only be applied in a \emph{first-person} sense \citep{adlamnonabsolute}: this involves maintaining that quantum mechanics always predicts the right correlations for any set of measurement outcomes which can be jointly witnessed by an individual observer, but its predictions are not  necessarily applicable for the correlations across pairs  of outcomes which can never be compared by anyone. This amounts to denying an assumption that  \cite{schmid2023review} call `Born Inaccessible Correlations,' which requires that when two measurements are made in parallel the two outcomes always arise with a joint frequency given by the Born rule, even if no single observer could ever access both outcomes. Once this is denied, it would seem that only the pair $(B,D)$ needs to obey the predictions of quantum mechanics, which can be achieved  without violating the Local Friendliness inequalities.

But we can imagine giving Bob a choice. Either  he performs the measurement to acquire the value $B$, or else he performs a  measurement  in a basis which is suited to establish the result of Alice's measurement, thus acquiring the value $A$ instead - for example, he can do this by simply going into  Alice's laboratory and asking  about her outcome. Divya can  be given a similar option to choose between acquiring the value $D$ or acquiring the value $C$ instead, and then she can tell Bob her outcome. This suggests that although only one of the pairs $\{ (B,D), (B, C), (A, D), (A, C) \}$ will be observed on each run, any one of these pairs \emph{could} be observed by a single person, Bob, on any given run. It follows that statistically, after many repetitions of the experiment, if we assume that quantum mehcanics is correct in a first-person sense then all four pairs must obey the quantum predictions over the runs where they are observed. Moreover ruling out retrocausality and superdeterminism  ensures that the outcomes obtained by Alice and Chidi can't depend on which of the four pairs Bob and Divya choose to access, and the locality assumption ensures that Bob's outcome can't depend on Divya's choice or vice versa. So it follows from the Local Friendliness assumptions that if the predictions of quantum mechanics are always right in a first-person sense, there must be a single fixed probability distribution over the four variables $A, B, C, D$ which reproduces  the pairwise predictions of quantum mechanics for \emph{all} the pairs $\{ (B,D), (B, C), (A, D), (A, C) \}$\footnote{For mathematical convenience it is common to represent the Bong et al experiment in terms of marginals over a probability distribution. However this obscures the important difference between first and third-person paradoxes, as discussed further in section \ref{EWF}, so in this article I will stick to the more physically transparent story in terms of one pair  actually being accessed at the end of the experiment.}.  The theorem of Bong et el proves definitively that this cannot be done, so it seems that if we want to maintain the first-person universality of quantum mechanics we must deny one of the Local Friendliness assumptions.

One popular proposal for resolving this paradox involves  denying `absoluteness of observed events,' i.e. the assumption that observed events are absolute single events, not relative to anything. If we deny absoluteness of observed events, it follows that there need not exist any well-defined probability distribution over the four outcomes $A, B, C, D$, in which case it is impossible to derive the Local Friendliness inequalities and hence no contradiction with the predictions of quantum mechanics can arise.  

However, non-absoluteness of this kind seems very problematic from the point of view of the epistemology of science \citep{https://doi.org/10.48550/arxiv.2203.16278,adlamnonabsolute}. This is because denying absoluteness of observed events implies that when Alice tells Bob about her outcome, the value he learns does not correspond to an absolute fact about what she actually observed prior to the conversation, and thus it commits us to postulating some kind of  instability in memory, communication or language itself such that each of us is trapped within a set of facts relativized to the self and/or the present moment, with no possible means of learning about any regularities existing outside of this one set of relative facts. This problem could perhaps be mitigated by putting forward some principled reason why  failures of memory, communication or language occur only in \emph{this} specific kind of case, but since the scenario where Bob asks Alice about her measurement outcome to acquire the value of $A$ is just an ordinary conversation between two people, it's unclear that there is any robust justification for the assertion that the epistemic difficulties  will remain limited to one specific EWF scenario. So there are quite strong reasons to think that absoluteness of observed events is an undesirable option for responding to the theorem,  thus motivating a search for  better alternatives.

\section{The reference system Loophole \label{loophole}}

In this section I will show that in addition to the three assumptions which together constitute `Local Friendliness,' there is an additional assumption required to arrive at the Local Friendliness inequalities, related to the role of reference systems. I will demonstrate that denying this assumption offers a `loophole' which can be used to retain Local Friendliness while still preserving the observable first-person predictions of quantum mechanics, in a sense to be further specified below.

In this section and the next  I will focus entirely on the case where the Bong et al experiment is performed using a pair of particles $S_A, S_C$ which are entangled in their spin orientation degree of freedom, so Alice's measurement on $S_A$ involves comparing the direction of its spin to a specific physical axis that she stores in her laboratory. This case offers a convenient illustration because it is straightforward to see why and how a variable like spin must be relativized to a physical reference axis; in section \ref{otherobservables}, I will discuss  whether the same approach can be taken for implementations of the Bong et al experiment using other kinds of variables. 
 
The reference system loophole arises from noting that there are two different things we could mean when we refer to `the outcome of Alice's measurement.' On the one hand this description could refer to what Alice actually sees during the measurement, i.e.  either  `the spin of $S_A$  is parallel to the axis in my laboratory,' or  `the  spin of $S_A$  is antiparallel to the axis in my laboratory.'  I will refer to this variable as $A_I$, assigning it values $\{ +1, -1\}$ corresponding to `parallel' and `antiparallel' respectively. Alternatively, `the outcome of Alice's measurement' could also refer to the variable stating the direction that Alice's spin is ultimately found to point \emph{relative to Bob's axis} in the external laboratory. I will refer to this variable as $A_E$, assigning it values $\{ +1, -1\}$ corresponding to `parallel' and `antiparallel' respectively. These two variables are distinct because Alice's axis and Bob's axis are different physical systems   so it is at least in principle possible that $A_I$ and $A_E$ might be different.

  For example, suppose that Alice performs her measurement and sees that the spin of $S_A$  is parallel to her axis. Assuming that Alice and Bob started out with their axes aligned, we would normally suppose that this means  the spin of $S_A$  is parallel to Bob's axis too. But suppose that when Bob opens Alice's laboratory and goes inside he discovers that Alice's axis (and with it, the rest of her laboratory) has flipped around, so it is now antiparallel to his axis. Then clearly the spin of $S_A$ is now  antiparallel relative to Bob's axis, so $A_I$ and $A_E$ do not necessarily have to be the same. 

 Thus let us define an additional variable  $R_{AB}$ which takes values $\{ +1, -1 \}$ such that $+1$ means `Alice's axis is found to be parallel to Bob's axis,' and $-1$ means `Alice's axis is found to be antiparallel to Bob's axis.'\footnote{For simplicity I will assume these are the only two possibilities, though in principle one could imagine models in which Alice's axis can sometimes be found in some other orientation relative to Bob's axis.}. So  $A_E = A_I \times R_{AB}$,  i.e. the value of $A_E$ depends on the results of two distinct measurements which fix values for $A_I$ and $R_{AB}$. The standard presentation of the Bong et al experiment assumes that if Alice and Bob initially share a reference system and set their axes in alignment then they will always remain aligned, i.e. it is always the case that $R_{AB}= +1$, but since $A_I$ and $A_E$ are distinct this does not necessarily have to be the case. In the rest of this section I will explain how allowing $A_I$ and $A_E$ to be different from one another can allow us to preserve the predictions of quantum mechanics without denying Local Friendliness, but later I will explain in more detail why one might think this is actually plausible.
 
As described above, the proof of \cite{Bong_2020} proceeds by arguing that if there are `absolute' facts about what everyone sees in this experiment and we assume locality and no-retrocausality, then we must be able to write down a well-defined probability distribution over the four measurement outcomes $A, B, C, D$, and if quantum mechanics is correct in a first-person sense then this distribution should reproduce the relevant pairwise predictions of quantum mechanics for  $\{ (B,D), (B, C), (A, D), (A, C) \}$. But we can now see that there is an ambiguity here, because $A$ could refer to either $A_I$ or $A_E$ (and likewise for $C, C_I, C_E$) - and in fact, whichever one we choose, the contradiction no longer arises. 

First suppose that we identify $A$ as $A_I$ and $C$ and $C_I$, so we are considering the four variables $A_I, B, C_I, D$. Since $A_I$ and $C_I$ describe what Alice and Chidi actually see, absoluteness of observed events indicates that they must have well-defined absolute values on every run of the experiment. Thus if we also assume locality and no-superdeterminism, there must be a well-defined probability distribution over the four variables $A_I, B, C_I, D$ which satisfies the Local Friendliness inequalities. However, $A_I$ is not defined relative to the same reference system as $C_I$ or $D$, and as I will explain in more detail in section \ref{superposition}, in fact neither standard quantum mechanics nor the QRF formalism makes any unambiguous predictions about what the correlations between $A_I$ and $C_I$ or $A_I$ and $D$ should be, particularly if we are only assuming first-person universality. Thus although Local Friendliness does imply that the distribution over $A_I, B, C_I, D$ must obey the Local Friendliness inequalities, there is no reason to expect that the pairwise correlations over  $\{ (B,D), (B, C_I), (A_I, D), (A_I, C_I) \}$ will match the quantum predictions that Bong et al use to arrive at a contradiction (or any other  set of quantum predictions), so no contradiction arises here. 

Next suppose that we identify $A$ as $A_E$ and $C$ as $C_E$, so we are considering the four variables $A_E, B, C_E, D$.  Now all four variables are defined relative to the same reference system, so we would expect the pairwise correlations over  $\{ (B,D), (B, C_E), (A_E, D), (A_E, C_E) \}$ to match the quantum predictions used by \cite{Bong_2020} to derive their contradiction. However, $A_E$ depends on both $A_I$ and $R_{AB}$, so it has a well-defined value only when there is a fact about the value of $R_{AB}$. In particular, if we suppose that $R_{AB}$ only comes to have a well-defined value if Bob actually measures it, it follows that $A_E$ will not have a well-defined value on runs of the experiment where Bob does not perform a measurement to determine the orientation of Alice's axis relative to his. Moreover, in order for Bob to perform the supermeasurement which yields a value for $B$ he has to keep Alice's laboratory closed and prevent any information about the outcome of Alice's measurement from escaping, and moreover all information about her outcome is subsequently erased in the measurement of $B$. I will assume here that this means that Bob cannot possibly measure both $B$ and $R_{AB}$ on the same run of the experiment (this assumption is explained further in section \ref{distinct}).  Therefore $A_E$ and $B$ will not ever have simultaneously well-defined values, in which case Local Friendliness does not imply that there should exist a single well-defined probability distribution over the variables $A_E, B, C_E, D$ or that it should obey the Local Friendliness inequalities, so  again  no contradiction arises. That is, regardless of whether we take $A = A_I$ or $A = A_E$ we find that there is no contradiction between Local Friendliness and the predictions of quantum mechanics, properly understood.

 \subsection{Toy Model \label{toy}}

 To give a clearer sense of how the reference system loophole might work, here is a simple toy model which achieves the desired statistics over a series of $N$ trials. This model is based on simply supposing that the results of Alice and Charlie's initial measurements are uniformly random and uncorrelated with any of the other variables in the experiment, so all the correlations are put in later. However  this is only one example and there could be many other ways to implement the loophole. 

 For each trial in the sequence:

\begin{itemize} 
\item  Generate a value for $A_I$  by selecting $+1$ or $-1$ uniformly at random. 

\item  Generate a value for $C_I$  by selecting $+1$ or $-1$ uniformly at random. 

\item Randomly select one of the pairs  $\{ (B,D), (B, C_E),$ $ (A_E, D), (A_E, C_E) \}$ to be accessed by Bob at the end of the experiment.

\item Generate outcomes for the pair of variables that Bob accesses by drawing them according to the  probability distribution predicted by quantum mechanics. Do not generate any values for $A_E$, $C_E$, $B$ or $D$ on trials where they are not accessed by Bob.

\item If $A_E$ is accessed by Bob on this trial, set $R_{AB} = A_E/A_I$. 

\item If $C_E$ is accessed by Bob on this trial, set $R_{CD} = C_E/C_I$. 
\end{itemize}

In this model  $A_I$ and $C_I$ have well-defined absolute values on every trial, uncorrelated with the subsequent choice about which pair of variables will be accessed. Additionally in each trial all the other variables which are actually observed have well-defined absolute values. So the Local Friendliness assumptions are always satisfied for all of the observed outcomes\footnote{The model is still nonlocal in the sense discussed in section \ref{locality}, but it satisfies the locality assumption which appears in the Bong et al proof.}. Yet by construction the two outcomes that Bob actually observes at the end of the experiment always obey the predictions of quantum mechanics. So this model indeed makes Local Friendliness compatible with the first-person universality of quantum mechanics, in a sense that will be discussed in more detail in section \ref{preservation}.

\subsection{State Description \label{statedescription}}

As described in section \ref{Wigner}, the usual formulation of the Wigner's Friend and Extended Wigner's Friend scenarios supposes that when Alice performs her measurement her state relative to Bob transitions as follows: 

\begin{equation}  x | \uparrow \rangle_S + y | \downarrow \rangle_S) \otimes | \psi \rangle_A \  \rightarrow \ x | \uparrow \rangle_S | see \ up \rangle_A + y | \downarrow \rangle_S  | see \ down \rangle_A ) \label{eq1} \end{equation}

The reference system loophole envisions  a different state transition. In this proposal there is always a single definite fact about what Alice saw, which is selected stochastically, like  in  a traditional wavefunction collapse. To be definite, I will consider the case where the  outcome of Alice's measurement is that she observed the particle to be parallel to her axis, i.e. $A_I = +1$. Then the transition will instead appear  as follows: 

\begin{equation}  x | \uparrow \rangle_S + y | \downarrow \rangle_S) \otimes | \psi \rangle_A \  \rightarrow  x | \uparrow \rangle_S | P \rangle_A  + y | \downarrow \rangle_S | P' \rangle_A )  \end{equation}

Here, $| P \rangle_A $ represents a  state of Alice where she has observed the particle to be parallel to her own axis; $| P' \rangle_A $  also represents a state of Alice where she has observed the particle to be parallel  to her own axis, but it has a different total orientation relative to Bob.  That is,  $| P \rangle_A$ and $|P' \rangle_A$ both have the same \emph{internal} structure,  but they  stand in different external relations to Bob. $|P \rangle_A$ and $|P' \rangle_A$ are identical from Alice's point of view as long as she has access only to facts about the inside of her lab, but they are not identical from Bob's point of view\footnote{ Another possible way of formalizing this proposal would be to treat `the facts about what Alice has seen' and `Alice's orientation with respect to Bob' as two distinct quantum degrees of freedom. Thus the post-measurement state relative to Bob, in the case where Alice gets the `parallel' outcome, would look something like$   x | \uparrow \rangle_S | \uparrow \rangle_A | P \rangle_A  + y | \downarrow \rangle_S  | \downarrow \rangle_A | P \rangle_A )$, where $| \uparrow \rangle_A$ and $| \downarrow \rangle_A$ correspond to different states for the orientation degree of freedom for Alice's laboratory, while $| P \rangle_A$ is the `parallel' state of the `what Alice has seen' degree of freedom. The reason I have not adopted this approach is because, as explained in section \ref{symmetry}, one natural interpretation of this proposal is that `what Alice has seen' should not be treated as a quantum degree of freedom at all, in which case it should not be represented as a quantum state in a Hilbert space.}.  

Note that this state  is still a superposition relative to Bob in the orientation degree of freedom, so if he performs an appropriate measurement he will be able to detect interference between the branches, just as quantum mechanics predicts. That is, the results of probing this state in the orientation basis will be no different from the results of probing the state $ x  | \uparrow \rangle_S | see \ up  \rangle_A + y | \downarrow \rangle_S  | see \ down \rangle_A $ as in the standard presentation of the Wigner's Friend experiment. But also, in both branches of the superposition Alice's internal structure encodes the result `parallel,' so when Bob performs a measurement to ascertain the \emph{internal} structure of her state (for example,  by simply asking her about her outcome) he will always learn that she saw `parallel,' regardless of whether or not her axis was found to be up or down relative to him. So we see that it is possible both for Alice and her laboratory to be in a superposition relative to Bob, and also for there to be a well-defined and communicable fact about what she has personally observed.

We can straightforwardly apply this approach to the Extended Wigner's Friend scenario of \cite{Bong_2020}, performed using spin measurements. Suppose that Alice and Chidi are measuring a pair of particles $S_A$, $S_C$ originally prepared in the state $ x  | \uparrow \rangle_{S_A} | \uparrow \rangle_{S_C}     + y | \downarrow \rangle_{S_A} | \downarrow \rangle_{S_C}    $. After Alice and Chidi have performed their measurements each of  them has a well-defined internal configuration which records a specific outcome of the  measurement, but they are still in a superposition in the orientation basis relative to the external frame occupied by Bob and Divya. Suppose for example that Alice saw `parallel' and Chidi saw `antiparallel.' Then the appropriate state is:  

\begin{equation} x  | \uparrow \rangle_{S_A} | \uparrow \rangle_{S_C} |P \rangle_A | A \rangle_C   + y | \downarrow \rangle_{S_A} |\downarrow \rangle_{S_C} |P' \rangle_A | A' \rangle_C    \label{eq3}  \end{equation}

As in the previous case, $| P \rangle_A $ represents a state of Alice where she has observed the particle to be parallel  to her own axis and and $| P' \rangle_A $ is a subjectively identical state with a different overall orientation; similarly for  $| A \rangle_C $ and $| A' \rangle_C $. It can be seen that in the orientation basis we have an ordinary quantum superposition relative to Bob and Divya, so if Bob and Divya perform their non-commuting measurements to obtain the values $B$ and $D$ by measuring the  appropriate orientation variables, they will see  results corresponding to the standard quantum predictions. 

But if Bob instead tries to find out the value $A$, there are two different measurements that must be disambiguated. First, Bob can ask Alice what the result of her measurement was, and presumably if he does this correctly he learns the value of $A_I$ as defined above. This variable has a definite well-defined value on every run of the experiment, but from the state in equation \ref{eq3} we can see that it  has the same value on both branches of the superposition, so the result of the measurement on $C$ or $D$ will not be  correlated with it\footnote{Note that I will argue later that `the result of Alice's measurement' is a complete observable and cannot be directly measured, so the process by which Bob finds out Alice's outcome is not merely a measurement on a quantum degree of freedom; see section \ref{moregeneral} for further discussion.}.

Alternatively Bob could perform a measurement which jointly probes Alice's outcome and also the direction of her axis relative to his own laboratory, corresponding to the variable $A_E$ as defined above. In this case he is in effect just performing a measurement of Alice's spin relative to his own reference system, and from the state in equation \ref{eq3} it is clear that this experiment will reproduce the expected quantum statistics.

Thus the state in equation \ref{eq3} is one in which the observed outcomes are absolute and yet  the standard quantum predictions for measurements in the orientation basis continue to hold, \emph{without} retrocausality or superdeterminism. That is, insofar as we agree that the variables to which the predictions of quantum mechanics are supposed to apply are $A_E, C_E$ and not $A_I, C_I$, this state reproduces all the standard predictions of quantum mechanics  without violating the Local Friendliness assumptions.

 \subsection{Objections}
Now  I will discuss a few possible  objections to this proposal.

\subsubsection{Preservation of quantum mechanics \label{preservation}}

I have suggested that the reference system loophole allows us to keep Local Friendliness without denying any of the predictions of quantum mechanics. However, one might naturally object that the predictions of quantum mechanics are preserved  only in quite a specific sense. 

Consider the case where Bob declines to measure $B$ and instead seeks to learn the value of $A$, so he goes into Alice's laboratory and asks her what result she saw. The answer to this question corresponds to the variable $A_I$. Separately Bob also learns $R_{AB}$, since attempting to enter Alice's laboratory will of course make it clear to him whether Alice's laboratory has reversed its orientation relative to him or not. He can thus calculate $A_E$ and compare its value to $C_E$ or $D$, and he will find that these correlations satisfy the quantum predictions for the relevant entangled state used to set up the experiment. 

But I have also suggested that the correlations between $A_I, C_E$ or $A_I, D$ will not generally conform to the quantum predictions for this entangled state. Moreover, I have specified that  Bob  learns $A_I$ and $R_{AB}$ separately, so he will be able to observe empirically that the correlations between $A_I, C_E$ or $A_I, D$ do not conform to the quantum predictions for this entangled state. So one might complain that Bob has in fact seen an empirical violation of the quantum mechanics after all.

However, that is simply to say that if Bob does the experiment wrong he will see the wrong results. As we will see in section \ref{superposition}, quantum mechanics (or at least standard first-person quantum mechanics)  does not predict anything in particular about the correlations between $A_I$ and $C_E$ or $D$, so it  isn't possible to observe a violation of the quantum predictions by looking at these correlations; if Bob wants to see the correlations predicted by quantum mechanics he needs to be looking at the right variable, which in this scenario is $A_E$. 

Now, \cite{dbdlh} criticize this proposal on the basis that is `a modification of the experimental protocol ... (because) the outcomes recorded by the superobservers in the LF statistics are not their friends’ observed results, but the orientations of the spins relative to their own laboratories.' But there is no modification here; this is simply a \emph{clarification}. In the original   proposal of \cite{Bong_2020} it is taken for granted that $A_I$ and $A_E$ are always the same, so the outcome $A$ as recorded in the LF statistics is assumed to represent \emph{both} Alice's observed result (that is why the absoluteness of observed events assumption is  applied to it) and \emph{also} the orientations of the spins relative to the external laboratories (that is why Bong et al compare the correlations involving $A$  directly to predictions derived from two-qubit photon polarization states all stated using the same polarization axis). Whereas if $A_I$ and $A_E$ are  not actually the same then it is simply impossible for a single variable to refer to both of these things at once, so we can only make sense of the experiment by deciding whether $A$ is supposed to refer to $A_I$ or $A_E$. Both are plausible interpretations of what is written in the Bong et al paper; and in any case  it does not matter which one we choose because  either way we no longer end up with a contradiction. 

Here, \cite{dbdlh} object: `The statistics collected as per Adlam’s proposal are then not the statistics the theorem bounds, and the fact that they saturate the Tsirelson bound carries no implication for local friendliness: the experiment performed is not an extended Wigner’s friend experiment in the sense of Bong et al.' But this is exactly the point of the loophole! The idea is precisely that Local Friendliness is relevant only to the four variables  $A_I, B, C_I, D$, whereas the predictions of quantum mechanics instead apply to the variables $A_E,B, C_E, D$, so if  these two sets of variables are  distinct it follows   that quantum mechanics does not predict any violations of Local Friendliness and thus there is no paradox to be defused.

Of course, one might still object that if Bob actually performs this experiment and observes empirically that $A_I$ and $A_E$ are different, he has in some sense found an empirical violation of quantum mechanics, because the divergence between $A_I$ and $A_E$ is a phenomenon which is not predicted by standard quantum mechanics. However, it is not that quantum mechanics makes the \emph{wrong} prediction here - rather it does not really predict anything in particular for $A_I$. So Bob has not disproved quantum mechanics, he has simply demonstrated that it is not a universal theory of everything, which we already had good reason to believe. But nonetheless the setup does  retain the superposition state of Alice and $S_A$ together and ensures that  relative to Bob this state continues to evolve unitarily until he measures it, so there is a robust sense in which the quantum predictions have been preserved as long as we are clear about the variables to which those predictions are supposed to apply.

\subsubsection{Distinct measurements \label{distinct}}

As described above, it is essential to the formulation of the reference system loophole that $B$ and $A_E$ never have simultaneous values. The justification for this is that in order to perform the supermeasurement to learn the value of $B$, Bob must erase all information about the result of Alice's measurement, so when Bob measures $B$ he cannot also measure $R_{AB}$ and hence $A_E$ does not come to have a well-defined value.

One might object to this on the basis that the result of Alice's measurement is $A_I$, and $R_{AB}$ is a distinct degree of freedom from $A_I$, so Bob ought to be able to measure $R_{AB}$ without measuring $A_I$. In particular, \cite{dbdlh} note that there are  various protocols in the literature where we imagine  an observer performing a measurement inside a closed laboratory and then sending some piece of information out of the laboratory while keeping the specific value of the measurement outcome confined to the laboratory \citep{Deutsch1985QuantumTA,ying2023relating}; it can be shown  that  as long as  the information sent out is uncorrelated with the  result of the internal measurement, it is still possible for an external observer to perform a supermeasurement. For example, it is possible to send out a message saying `the measurement was performed and had an outcome' and then still perform a supermeasurement, provided that we don't say what the outcome was. So one might think it would similarly be possible to send information about $R_{AB}$ out of the laboratory without preventing Bob from measuring $B$. 

However, this is not at all the same as the scenario envisioned in the reference system loophole, because in that case the variable $R_{AB}$ is an external one pertaining to the relation between Alice and Bob, so it certainly cannot be measured by Alice within her laboratory and then sent out in a message. Furthermore, ex hypothesi in this setup  the outcome relevant to the quantum statistics is $A_E$, not $A_I$. What this means is that Bob will be unable to measure $B$ if he gets any information about the variable $A_E$, because it is $A_E$ and not $A_I$ which the role of `the result of the internal measurement' in the protocols of \citep{Deutsch1985QuantumTA,ying2023relating}. Thus since  $R_{AB}$ is a part of $A_E$, it  follows  Bob cannot measure $B$ if he has previously measured $R_{AB}$, since in that case he has obtained some information about the variable $A_E$.

The analysis of \cite{dbdlh} ultimately backs this up: they note that in the scenario relevant to the reference system loophole, `the orientation of Friend’s frame becomes entangled with the spin and, as a consequence, sharing information about Friend’s orientation would decohere the superposition of Friend’s lab and spin.' So it seems evident that under the hypothesis considered here, Bob cannot ever measure both $R_{AB}$ and also $B$,  so it is indeed true that $A_E$ and $B$ are never simultaneously defined\footnote{\cite{dbdlh} go on to give a calculation showing that if we assume the orientation and spin do not become entangled it's possible to send information about the orientation out of the laboratory without preventing Bob from doing his measurement later - but of course this is irrelevant to the hypothesis under consideration, where they \emph{do} become entangled.}.

\subsubsection{ Maintaining Alignment} 

Another natural worry is about how exactly Alice ends up losing her alignment with Bob. Of course Alice and Bob have  a well-defined orientation relative to each other prior to the experiment - that is, prior to the experiment they would presumably set up a pair of axes and compare them in order to agree the direction in which they plan to measure the spins. And it seems as though nothing particularly disruptive happens during the experiment to alter that, so shouldn't we expect to find that their axes are still aligned at the end, such that  $R_{AB}$ will always be equal to $ 1$?

However, it must be emphasized that the conditions for Bob to actually be able to perform the supermeasurement  are very stringent; no information about Alice's outcome can be allowed to escape from the laboratory, which in practice means that her laboratory must be completely isolated from all interactions with the outside world, including gravitational interactions. We have no empirical experience with laboratories which are \emph{completely} cut off from the external world and kept in coherent quantum superposition in this way, so we cannot necessarily expect that our intuitions about what would happen in this scenario are a reliable guide. 

Of course we probably do not want to say that merely the act of isolating Alice's laboratory causes her to lose her alignment with Bob, since we know that  small quantum systems can be prepared in  a pure state of well-defined alignment relative to a laboratory frame and then in isolation they will remain in the same pure state relative to the laboratory for apparently arbitrary lengths of time. But the combination of  complete isolation \emph{plus a quantum measurement performed inside the isolated system} is not something that we can test empirically at the moment, so it is consistent with everything we currently know to maintain that Alice loses her well-defined orientation relative to Bob as a result of performing her measurement on the qubit $S$. And indeed, this makes sense: by definition $S$ does not have a definite orientation relative to Bob's reference system, and in the process of performing the measurement Alice aligns her own laboratory with $S$, so if we maintain that Alice  always sees a single definite outcome of her measurement and that the superposition does not collapse relative to Bob, it immediately follows that in the course of performing the measurement she \emph{must} cease to have a definite orientation relative to Bob's reference system. 

One might  worry that this involves a puzzling  asymmetry: why should the interaction between $S$ and Alice result in Alice's reference system going out of alignment with Bob's reference system, rather than $S$ being dragged into alignment with Bob's reference system? The latter might seem more natural, given that Alice's laboratory frame is so much larger than $S$. And if such a thing were to happen it would  simply be a collapse of the wavefunction from Bob's point of view, after which he would not be able to detect interference in the orientation basis. 

 In fact, the reason for this asymmetry  is simply that in the model considered here this is just what it \emph{means} for Alice and $S$ to undergo unitary evolution relative to Bob. The process of `going out of alignment' is really the same process that occurs in the usual quantum description of the Wigner's Friend scenario as laid out in equation \ref{eq1}, where Alice and the reference system end up in an entangled superposition - the only difference is that instead of   \emph{all} of Alice's degrees of freedom becoming superposed relative to Bob, in the scenario relevant to the reference system loophole  only some of her degrees of freedom become superposed. If you consider the standard quantum description of a Wigner's Friend experiment plausible - that is, you think it is plausible that in such an experiment  Alice would go into superposition relative to Bob rather than the particle simply collapsing out of superposition - then you should also consider it plausible that only Alice's orientation degree of freedom becomes superposed relative to Bob, because that is a strictly weaker claim.

So in fact,  the question of  Alice going out of alignment with Bob versus the spin particle coming into alignment with Bob ultimately depends on Bob's  technical capabilities. Given the setup described above, if he is able to keep the quantum degrees of freedom sufficiently isolated to ensure that they continue to evolve unitarily relative to him, then Alice necessarily goes out of alignment with Bob, because otherwise (given that her measurement always produces a definite outcome) the evolution would not be unitary. Whereas if Bob is not able to keep the quantum degrees of freedom well-isolated, then the particle comes into alignment with Bob instead, and it is no longer possible for him to see the usual quantum correlations when he measures Alice and her laboratory. This is really just the same as the standard account of the experiment: if the isolation is not good enough then decoherence ensures that the wavefunction effectively collapses relative to Bob.

 \cite{dbdlh} object to the proposal on the grounds that `In practice, such a shared frame can be fixed and continuously monitored by reference to a shared external field, such as the gravitational field of the Earth, or some pre-aligned gyroscopes.' And of course  there are possible ways to maintain a shared frame - by default the whole community of human observers maintains a joint shared quasi-classical frame all the time! However, active monitoring to maintain a shared frame necessarily involves information about the variable $R_{AB}$  leaving Alice's laboratory, since Alice's orientation degrees of freedom must be in continuous interaction with the relevant external field or similar. And if we agree that as discussed in section \ref{distinct}  the variable $R_{AB}$ is a part of Alice's outcome, we know that Bob will not be able to perform the supermeasurement to establish the value of $B$ if any information about $R_{AB}$ gets out of the lab. That is, one of the consequences of the model set out here is precisely that if Bob wants to perform the EWF experiment and see correlations in accordance with the quantum predictions, he has to deliberately make sure that  a shared frame is \emph{not} maintained, so steps must be taken to block the continuous monitoring described by  \cite{dbdlh}.

\subsubsection{Predictions for other orientations \label{superposition}}

The  predictions invoked by \cite{Bong_2020} to argue that Local Friendliness predicts a violation of quantum mechanics are  for a set of spins all measured in the same reference system. Since the reference system loophole entails that  $A_I, B, C_I, D$ arise from measurements which are not all performed relative to the same reference system, it straightforwardly allows us to deny that the probability distribution over $A_I, B, C_I, D$ should be expected to match those particular probabilities, which defuses the version of the paradox presented in the original paper.

However, quantum mechanics can also make predictions for measurements performed relative to different frames, so one might think that we can just derive a different set of quantum predictions for  $A_I, B, C_I, D$, using either  standard quantum mechanics or perhaps  the QRF formalism \citep{2020acop,delahamette2021perspectiveneutral,giacomini2021quantum,2021qrft}. And one might imagine that we will then find that quantum mechanics does in fact predict a violation of the Local Friendliness inequalities across the variables  $A_I, B, C_I, D$, just in some different way.
 
 First, of course standard quantum mechanics can easily make predictions for  the scenario where Alice's frame and Bob's frame always  stand in a well-defined classical relationship to one another, since that just amounts to a change of basis. But this case is irrelevant since it is not the scenario employed in  the reference system loophole: as explained in section \ref{statedescription},  the reference system loophole presumes that the outcome of the measurement of $R_{AB}$ is \emph{stochastic}, i.e. the orientation of Alice's frame relative to Bob's frame is not always found to be the same, so we cannot simply apply the predictions of quantum mechanics for a rotated frame.

Second, as exhibited  by \cite{dbdlh}, the QRF formalism can make predictions for the scenario where prior to Alice's measurement her frame is in a coherent superposition relative to Bob's frame.  However, this case is also not the scenario employed in the reference system loophole. 
As laid out in section \ref{statedescription} Alice's reference system is not in a superposition prior to her measurement -  it enters a superposition \emph{during} the measurement, as a \emph{consequence} of the measurement having been performed, and in a way that depends on the outcome of the measurement. This scenario cannot easily be modelled in the QRF formalism, because that formalism does not  model measurements as having single definite outcomes at all.

Additionally, even if we ignore that point, the calculation of \cite{dbdlh} does not appropriately model the setup envisioned in the reference system loophole. To see this, let is distinguish between two possible ways of interpreting the calculation. First of all, we can consider the case that   $A_I$ and $R_{AB}$ are to be understood as distinct and independent degrees of freedom. However, unless $R_{AB}$ is always the same, this will produce incorrect predictions for the correlations between $A_E, D$. The simplest example of this is given by the scenario in which we choose an entangled state and a pair of measurements such that quantum mechanics predicts that $A_E$ and $D$ should be perfectly correlated or anticorrelated. Then  note that regardless of what correlation we specify for $A_I, D$, if $R_{AB}$ is independent from $A_I$ and not always the same, multiplying $A_I$ with $R_{AB}$ adds some extra stochastic noise which is not correlated with $A_I$  and   therefore ensures that $A_E$ and $ D$ cannot possibly be perfectly correlated or anticorrelated. Since the basic starting point of the reference system loophole  that $R_{AB}$ is not always the same and that  $A_E, D$ are the variables to which the predictions of quantum mechanics  apply, this cannot be the right model.

Alternatively, perhaps the calculation of \cite{dbdlh} should be interpreted as positing that $R_{AB}$ does depend on $A_I, D$. But the problem then is that in order to be able to reproduce perfect correlations and anticorrelations we must allow that $R_{AB}$ may be simply a deterministic function of $A_I$ and $D$ which simply flips the value of $A_I$ whenever necessary to ensure that the correlations between $A_E, D$ turn out right. And once we allow dependence of this general form, then in fact we can have \emph{any correlations at all} between $A_I$ and $D$ and still end up with the right correlations between $A_E$ and $D$.  This is exactly how the toy model in section \ref{toy} works -  $A_I$ can be completely uncorrelated with $D$ and then we can just use $R_{AB}$ to fix things retrospectively so $A_E$ and $D$ end up correlated the right way. Thus once we allow $R_{AB}$ to depend on $A_I, D$ in this way there is no reason for the quantum  formalism to place any particular constraints on the correlations between $A_I$ and $D$, because all the work to match the quantum correlations can be done by $R_{AB}$ instead of $A_I$.
 
 The reason  \cite{dbdlh} arrive at a definitive answer in their calculation is because they simply make a specific choice for a quantum operator defined in Alice's frame that they take to represent $A_I$ and then they use a relativization map to rewrite it as a frame-independent relational observable. But since the relativization map is merely a coordinate transformation, this method requires that  before the calculation even gets started they have already decided how the correlations with $D$ are distributed between $A_I$ and $R_{AB}$, and  the calculation just amounts to  rewriting those preselected correlations in a different way. Whereas the point of the reference system loophole is that there are \emph{many different possibilities} for how the measurement of $A_I$ could be correlated with $D$ which will still produce the right statistics from within Alice's local point of view and which will give the correct quantum correlations for $A_E, D$ in the end, provided that we use $R_{AB}$ appropriately to compensate for any deviations. This means that there are many different frame-independent relational observables which could potentially represent $A_I$, depending on how we expect the entanglement correlations to be distributed between $A_I$ and $R_{AB}$ - for example, instead of the relational observable calculated by  \cite{dbdlh} we could instead propose a relational observable which is completely uncorrelated with the relational observable associated with $D$, such that all relevant correlations between $A$ and $D$ appear in the measurement of $R_{AB}$ instead\footnote{It is also possible that $A_I$ simply cannot be represented by a quantum operator at all, an option that will be explored further in section \ref{symmetry}.}.  
 
 Of course we do not have any empirical evidence about how the relevant correlations should be distributed between $A_I$ and $R_{AB}$, since we have not performed any experiments where observers are superposed relative to each other, and the QRF formalism by itself cannot possibly  answer this question because the only thing it does it tell us  how to translate operators between frames once we have already decided which operators to use. So as a matter of fact neither standard QM nor the QRF formalism predicts anything specific for the correlations between $A_I$ and $D$, because neither one can conclusively determine how the relevant correlations ought to be split between $A_I$ and $R_{AB}$.

Finally, even if it were possible to get a sensible answer for the correlation between $A_I$ and $D$ out of the QRF formalism, we obviously should not consider such a prediction binding in the context of the Bong et al paradox specifically. This is because  the whole point of the Wigner's Friend scenario is that is \emph{mathematically impossible} that the predictions of ordinary quantum mechanics are correct across all four pairs $\{ (B,D), (B, C),$ $ (A, D), (A, C) \}$ simultaneously. That is, we already know for sure that in this kind of setup, even the ordinary predictions of standard quantum mechanics  for the correlations between measurements performed within a single reference system must fail to be upheld at some point\footnote{This is true even in the Everett interpretation and other approaches which violate absoluteness of observed events - in  that case  there is simply no well-defined joint distribution across certain pairs of observations, so the predictions of quantum mechanics cannot possibly be upheld across those observations, though it also would not be correct to say they are violated across those pairs: in effect such approaches are simply denying the law of excluded middle, allowing us to say that the predictions are neither correct nor incorrect.}. So the upshot is that  we are definitely going to have to accept some kind of failure or inapplicability of standard quantum mechanics to at least one of these four pairs, and  we have to decide   which mode of failure we find most palatable. In particular, the reference system loophole falls into the class of responses to the paradox aiming to maintain the first-person universality of quantum mechanics, meaning that the goal is to arrive at a sensible story in which the results across the pair of variables out of $\{ (B,D), (B, C_E),$ $ (A_E, D), (A_E, C_E) \}$ observed at the end of the experiment always obey the quantum predictions, while some of the quantum predictions across the pairs not directly observed fail to be upheld.  So, given that we are not even assuming the correctness of the predictions of \emph{standard quantum mechanics} across all pairs of measurement outcomes in this setup, it would make no sense to insist on the universal correctness of the predictions of a more speculative framework such as the QRF formalism, particularly given that  we have no  empirical evidence to support such  predictions.

\subsubsection{Locality \label{locality}}

I have argued that the reference system loophole  allows us to maintain the assumption of locality that goes into the Local Friendliness theorem. However, one might object that this loophole does in fact involve a form of nonlocality. This can be seen from the toy model in section \ref{toy}, which relies on generating predictions for $\{ (B,D), (B, C_E), (A_E, D), (A_E, C_E) \}$ according to the usual quantum predictions. Since each of these pairs corresponds to measurements on a pair of entangled systems, the results are of course nonlocal in just the same way that  measurements on entangled systems are typically nonlocal. 

However, the key point here is that the locality assumption used to prove the Local Friendliness theorem is supposed to be \emph{weaker} than the kind of locality involved in ordinary measurements on entangled systems, which means that  violating the locality assumption entering into the Local Friendliness theorem requires us to accept a stronger form of nonlocality than the type that appears in the ordinary Bell experiment. More specifically, the locality assumption used in the Local Friendliness proof is \emph{parameter independence}, which means that the choice of a parameter on the Alice/Bob side of the experiment does not influence any outcomes on the Chidi/Divya side of the experiment. Thus if we want to respond to the EWF experiment by rejecting the locality assumption, we are forced to reject parameter independence specifically. By contrast, Bell's theorem uses a locality assumption which is the conjunction of parameter independence and also \emph{outcome independence}, which means that the outcome on one side of the experiment does not influence the outcome on the other side \citep{Jarrett,Shimony,MYRVOLD2002435}. Thus in order to respond to Bell's theorem by rejecting the locality assumption, we can potentially choose to reject outcome independence rather than parameter independence; and to many people this looks like a more appealing route, since violations of parameter independence appear to involve some kind of nonlocal causal influence whereas violations of outcome independence just involve some kind of reciprocal nonlocal coordination \citep{Jarrett,Shimony,MYRVOLD2002435}. 

With this in mind, note that the nonlocality involved in the toy model is simply the same kind of nonlocality involved in any pair of measurements on a pair of entangled particles, which means that this type of nonlocality can be accounted for by violating outcome independence rather than parameter independence. Obviously this result will not be satisfactory to someone who thinks all forms of nonlocality must be avoided, but the key point is that we no longer need any nonlocality beyond what we already had in an ordinary Bell experiment, so by using the reference system loophole we can avoid being pushed into any stronger form of nonlocality. For example,  if one already has in mind some model for getting rid of nonlocality in the Bell experiment - e.g. by appeal to retrocausality \citep{RevModPhys.92.021002} -  the reference system loophole ensures that this model can be made to work  exactly the same way in the EWF experiment, so there is no need to make a new model to accommodate a stronger form of nonlocality.

\subsubsection{Conservation Rules} 

Adopting the reference system loophole entails that there is some probability that when Bob opens Alice's laboratory after the experiment he finds that the entire laboratory has the opposite orientation from the way it started; so from Bob's point of view it looks  as if Alice's laboratory has `jumped' from one orientation to another. One might worry that this will violate various conservation rules, e.g. the conservation of angular momentum.

However, conservation rules hold \emph{within} a given frame of reference, not across different frames of reference. Given a system with a certain fixed momentum  described relative to one frame of reference, that system will have a different momentum  when we describe it relative to another frame of reference, but the change in its momentum is not a violation of the conservation of momentum; that rule holds only relative to any individual frame of reference. Thus since the phenomenon described here involves an interaction between several distinct frames of reference, we need to be careful about how exactly  conservation rules apply.

It might at first appear that the scenario suggested by the reference system loophole does involve a violation of conservation rules relative to Bob's reference system, for if Alice's laboratory starts out `the right way up' relative to Bob's laboratory, and then later it is found to be the other way up relative to Bob's laboratory, it seems that the angular momentum and energy needed to make the jump just appear from nowhere, so conservation of momentum and energy are violated relative to Bob's laboratory. 

However, as a matter of fact there is no extra energy or momentum involved at any stage: prior to the experiment Alice's laboratory is at rest relative to Bob's laboratory, and afterwards Alice's laboratory is also at rest relative to Bob's laboratory, just in a different orientation, so the total energy and momentum is the same\footnote{In order to make Bob's supermeasurement possible  Alice's laboratory must be shielded from an relevant graviational interactions which might induce decoherence, so we can safely assume that the potential energy is the same in both arrangements}. In particular, one should not make the mistake of imagining that there is an intermediate time at which Alice's laboratory is in motion relative to Bob's laboratory: rather what happens is that Alice's laboratory simply ceases to have any well-defined state at all relative to Bob's laboratory, and then later it has a well-defined state again, one which has the same total momentum and energy. We should not think of the `jump' as a literal physical jump which requires momentum and energy. 

In fact, the `jump' that Alice's laboratory performs is no different from what happens to any quantum system when it is measured. For example, imagine starting from  a particle in position $x$,  then performing  an operation which puts it in a superposition of position $x$ and position $y$, then measuring it and finding it in position $y$. It looks as though the particle has `jumped' from $x$ to $y$, which might seem like a violation of conservation of energy or momentum. But this is not usually considered a real violation of the conservation rules, because the jump is not a literal physical jump and we are never going to `catch' the particle in motion between $x$ and $y$: in the end it just ends up in a different position with the same energy and momentum. Of course this phenomenon seems more startling when illustrated with a macroscopic  system, but if we really believe that quantum mechanics applies to macroscopic systems as well as microscopic systems then there is no difference in principle between the two cases  -   we should not have double standards in the way we judge macroscopic quantum phenomena as compared to microscopic quantum phenomena.

\section{Symmetries\label{symmetry}}

Having demonstrated that the reference system loophole  is at least formally capable of providing a response to the EWF no-go theorem, in this section I will turn to the question of why one might think this is a \emph{plausible} response. In particular, I will argue that the proposed loophole fits in naturally with various facts about the symmetries of the relevant systems, and that it is  linked with ongoing research in relativistic physics and quantum gravity in ways which make this route seem  worth pursuing further. In this section I will continue to focus exclusively on the case of spin, and I will consider the generalization to other types of variables in section \ref{otherobservables}.

Recall that the reference system loophole arises from supposing that at the time of Alice's measurement there is an absolute fact about what Alice has observed, but there is not an absolute fact about the orientation of Alice's reference system relative to Bob's. The reason this is a natural thing to do  is because `orientation' is linked with a \emph{symmetry} of the laboratory subsystem. 

A theory is said to have a symmetry if there exists a class of transformations $\{ T\}$ which leave its equations of motion and solutions unchanged, meaning that the effects of such transformations cannot be detected. For example, in most modern spacetime theories `global translations' and `global rotations,' where we displace or rotate the whole contents of the universe in the same way, are symmetries. Usually the presence of a symmetry is interpreted as indicating that the relevant  transformations $\{ T\}$ do not produce  any  physically real change, which implies that any variable $V$ whose value can be changed by some transformation in this set is not physically real by itself.  

But we can also imagine applying transformations to proper subsystems of the universe. Suppose that $\{T \}$ is a symmetry group of the relevant theory, so applying a transformation in $\{ T \}$ to a proper subsystem doesn't change the equations of motion or solutions for that subsystem, and thus such a transformation cannot be detected from the point of view of an observer confined to that subsystem. For example, if we rotate Alice's whole laboratory and its contents all together, this change is not detectable from within the laboratory\footnote{I assume here that the laboratory is shielded in some way from the effects of gravity and other such external features which might make different orientations of the laboratory distinguishable; this is necessary because Bob will not be able to perform the supermeasurement to establish the value of $B$ if the laboratory is able to interact with the environment and undergo decoherence.}. However, such a transformation may still be detectable when we compare the subsystem to its environment \citep{greaves2011empiricalconsequencessymmetries}. For example, rotating Alice's laboratory changes the relation of the laboratory to the external world, and clearly this is a physically meaningful difference since it will have substantive empirical consequences when we open the door of the laboratory and Alice tries to exit. 

So if there is some variable $V$ of a proper subsystem which is changed by the symmetry transformations $\{ T \}$, the value of $V$ is not meaningful when we consider the subsystem in isolation, but it can still be meaningful when the subsystem is compared to an external reference system - that is, properties of this kind must be relativized to a reference system in order to have physical content. The QRF approach relies heavily on this way of thinking about the role of reference systems: `Note first that every reference system is associated with a symmetry group. For instance, a Cartesian frame is associated with the group of rotations SU(2), a clock (phase reference) is associated with U(1), and a reference ordering (which we shall consider in section III F) is associated with the symmetric group SN 6. If Eve does not share Alice and Bob’s RF, then she is ignorant of which element of the group describes the relation between her local RF and that of Alice and Bob' \citep{PhysRevA.70.032307}.  

This suggests the following conjecture about the relation between definiteness and symmetries, which can be used to give the reference system loophole a more principled formulation. Let us suppose that quantum degrees of freedom are always quantities which vary under symmetries and thus require relativization  to an external reference system. That is, let us suppose that invariant facts   are always definite and absolute, but properties which can be changed by symmetry transformations may enter into quantum superposition relative to some external reference system. In particular, if $V$ is a property of the subsystem   $S$  whose value can be changed by some kind of  symmetry transformations performed on $S$, then the value of $V$ may be   indefinite relative to an external reference $U$, even if there is nothing indefinite about the system $S$ in and of itself.

This conjecture gives us exactly what we need in order to formulate the reference system loophole and respond to the Local Friendliness theorem. It is natural to suppose that our actual measurement outcomes and observations must supervene on invariant facts (this assumption is standard in the physics literature  \citep{Loveridge_2018}) and thus  if the internal properties and relations of Alice and her laboratory  which are invariant under symmetry transformations  must always be well-defined, it follows that facts  such as `Alice found the spin of the particle to be parallel to her axis' will  always be well-defined, i.e. absoluteness of observed events will always be always upheld. But  after Alice performs a measurement on the orientation of the spin system, the resulting state can be written as a superposition such that in each branch of the superposition Alice's laboratory as a whole has a different orientation with respect to Bob's laboratory, but nonetheless the variables of her laboratory which live in the invariant subspace of the rotation group are the same in both branches, as shown in the state given in section \ref{statedescription}: 

\begin{equation} \psi_{SA}' =  a | \uparrow \rangle_S | P \rangle_A  + b | \downarrow \rangle_S | P' \rangle_A   \end{equation}

 So we can maintain that there is always a definite fact about what Alice observed and she always has a single definite conscious experience, even though her laboratory is in a superposition of two different orientations relative to Bob and thus Bob can still see results matching the quantum predictions for entangled particles.

Now there is a natural worry that one might have about the conjecture set out above. In general for any variable $V$ of $S$ which can be changed by symmetry transformations on $S$, we can define a corresponding invariant relation $R$ by relativizing that property to a reference system $U$ - for  example, the direction of spin can be turned into an invariant relative orientation between the spin system and the reference axis. But if the value of $V$ relative to $U$ is in a superposition, then one might think the invariant relation $R$ between $S$ and $U$ would also have to be in a superposition,  so  we cannot consistently maintain that the invariant relations are always definite while the corresponding relativized properties can be indefinite. 

But as a matter of fact, as shown by  \cite{Loveridge_2018} it is possible for a symmetry-variant property $V$ of $S$ to be indefinite relative to $U$ even though the state of $S \otimes U$ together does not involve any coherent superposition. I refer to \cite{Loveridge_2018} for more details, but the essential point is that when we relativize a variable $V$ to a reference system we do not simply relativize the variable to the system $U$ by itself: rather it must be relativize to a \emph{reference state} of the system $U$.  Thus even if the relation between $S$ and $U$ is always definite, given certain choices of reference state for $U$ the state of $S$ relative to $U$ can turn out to be a coherent superposition\footnote{In particular, if the reference state is a coherent superposition the state of $S$ relative to $U$ will also be a coherent superposition.}. I  defer further discussion of the details of this intriguing phenomenon to future work \citep{symmetriesref}, but the fact that such a thing is possible is an encouraging sign that the reference system loophole has some foundations in our existing understanding of relativization in quantum mechanics.

\section{Other Variables \label{otherobservables}}

The resolution of the paradox proposed above relies essentially on the fact that we are performing the Bong et al experiment using  a variable such as spin which varies under a symmetry transformation, i.e. the rotation group. But of course in principle the same experiment can be done with any variable  as long as it can be superposed and then measured by Alice. Will the same resolution work with other kinds of variables? 

\subsection{Symmetry-variant Quantities}

We can see that the reference system loophole will work for any variable which can be altered by a global symmetry transformation adequately similar to the rotation group. For example, it will also work with position, since position can be altered by global spatial translations and such translations are a symmetry of our existing physical theories. 

Imagine a version of the Bong et al experiment using superpositions in the position basis: Alice has a particle $S_A$ in her lab prepared such that in one branch of the wavefunction the particle goes down the `left' path and in another branch the particle goes down the `right' path, with both paths  defined relative to the external reference system occupied by Bob. Then let us suppose that at the time when Alice is still aligned with Bob's frame she  puts a red detector into the left path and a blue detector into the right path. In accordance with the spin case, let us suppose that on any given run Alice's measurement has one definite outcome, so the particle is found in some particular detector - e.g. let us say that it is found in the red detector. In this case, the way the reference system loophole works is that when Alice performs this measurement both she and her detectors lose their alignment with the external spatial reference system, and thus since `left' and `right' were defined relative to the external frame, there is no longer a well-defined fact of the matter about whether the red detector is in the left path or the right path relative to the external frame.  That is, from Bob's point of view Alice is now in a superposition of one branch in which the red detector is on the left path, and another branch in which the red detector is on the right path, with these two states corresponding to different overall positions of Alice's whole laboratory relative to Bob's laboratory:

 \[  \frac{1}{\sqrt{2}}  (| \uparrow \rangle_S | R \rangle_A  + | \downarrow \rangle_S | R' \rangle_A )  \]

Here, $|R \rangle$ and $| R' \rangle$ both represent  states in which the particle is found in the red detector, but from Bob's point of view they differ by a spatial translation. So just as in the spin case, the state shown above is a superposition relative to Bob, but nonetheless it represents a definite fact about the outcome that Alice saw and her corresponding conscious experiences. 
 
  Similarly, suppose we instead prepare a state for the system $S_A$ which is an energy basis superposition, i.e. $S_A$ has different energy levels in the two branches. Since energy is meaningful only relative to a standard of zero energy, there is  a symmetry corresponding to changing the zero of energy, so  we can imagine that this results in a superposition of two states of Alice's laboratory which internally look the same but which differ in their overall energy with respect to the standard of zero energy defined by Bob's external reference system. Thus again we can get a state which is a superposition relative to Bob but which represents a definite fact about what Alice actually observed. 

 \subsection{More General Superpositions \label{moregeneral}}

The reference system loophole will not work if the Bong et al experiment is performed using a superposition of the values of a variable   $V$ which is  invariant under all symmetry transformations. However, on reflection it is not obvious that we can actually perform the experiment using a superposition of this kind.

Initially it might appear that there exist superpositions of invariant quantities that we could use in such an experiment.  For example, imagine that both $S_A$ and $S_C$ are both composite systems of two particles, and each of them is prepared in a superposition of two different values for the relative distance between the two particles. One might at first think that relative distance is invariant under all relevant symmetry transformations, and of course we know  that superpositions of different values of relative distance are  possible - for example, the ground state wavefunction for the hydrogen atom,  $\psi(r) = \frac{e^{r/a_o}}{\sqrt{\pi}a_0^{3/2}}$ \citep{10.1088/978-0-7503-1167-0ch12} is a coherent superposition of different values for the relative distance $r$ between the nucleus and the electron. So there would seem to be no particular obstacle to performing the Bong et al experiment this this way. 

However, it is not really true that relative distance is invariant under all symmetry transformations. Of course it is invariant under spatial translations and rotations, but we must also consider time translations, which are a symmetry of most serious physical theories. To see this, note that it makes no sense to simply state a relative distance between two systems $X$ and $Y$, because relative distances can change, so the relevant variable is really `relative distance between $X$ and $Y$ \emph{at time $t$}.' Clearly this variable will be changed by time translations, so relative distance is not in fact invariant under all symmetry transformations: just like orientation and position, relative distance must be relativized to a `reference system,' which in this case will be a clock with a time reading. 

How would this work in the context of the reference system loophole? Well, in this scenario the role of the variable $A_I$ is played by `the relative distance between $A$ and $B$ observed by Alice when her laboratory clock reads time $t$' and the role of the variable $A_E$ is played by `the relative distance between $A$ and $B$ when Bob's laboratory clock reads time $t$.' When Alice performs  a measurement on the superposition of two different relative distances, she  sees that the relative distance between $A$ and $B$ is $x$  at the time when her clock reads $t$, so there is a definite, well-defined value for $A_I$. But similar to the spin case, we can imagine that Alice's laboratory clock  ceases to be synchronised with Bob's clock when she performs this experiment. So as in the spin case, if Bob simply asks Alice what she saw he will learn the well-defined fact about $A_I$, but in order to establish $A_E$ he also has to check whether their clocks are synchronised, so he may learn that Alice's clock read $t$ at the time when his own clock read some other time $t'$. Thus he will still be able to probe the superposition and see entanglement in the relevant quantum-mechanical variables $A_E$, even though he won't see these effects for $A_I$ on its own.  

More generally, it seems plausible that in the long run the class of relevant symmetries should include diffeomorphisms, since diffeomorphisms are a symmetry of general relativity\footnote{Here one might object that in some theories of quantum gravity spacetime is emergent so the diffeomorphism symmetry is only emergent, so it will not belong to the class of relevant symmetries in the end. However, of course that is true of any spacetime symmetry whatsoever, and the fact that the symmetry is only emergent does not change any of the considerations described here. Translation symmetry is only emergent because without spacetime there are no translations, but that does not mean that at a fundamental level absolute positions become meaningful without being relativized to a reference system; rather it simply means that at a fundamental level there is no such thing as position in either an absolute or relative sense.}. And once we include diffeomorphisms as a symmetry, there is a general argument which suggests that \emph{any} variable that we can directly interact with is necessarily symmetry-variant.

To see this, recall that in the context of a theory like general relativity which is subject to a diffeomorphism symmetry, \cite{Rovelli_2002} proposes that we should distinguish between `partial observables,' which can be changed by  diffeomorphisms, and `complete observables,' which cannot. Partial observables are typically quantities assigned to specific spacetime locations, while complete observables are global or relational quantities. Because partial observables vary under diffeomorphisms, the usual way of interpreting symmetry transformations indicates  that they are not physically real by themselves, but they can be meaningful when appropriately relativized to something. Moreover,  \cite{Rovelli_2002} suggests that  partial observables are physical quantities `to which we can associate a measuring procedure leading to a number,' while complete observables are  physically real but they are not the kind of thing we directly measure. That is, everything we interact with directly and couple our measuring devices to is a \emph{partial} observable, but the results of our measurements correspond to complete observables, because in the process of measurement the relevant partial observable becomes relativized to the measuring instrument and so the result of the measurement is in effect an encoding of that complete observable.

If we accept the idea that everything we measure directly is a partial observable while the result is a complete observable, it follows that  any variable that we can  measure directly must vary at least under diffeomorphism transformations. And of course the Bong et al experiment can only be performed using a superposition of variables that Alice is able to measure, so it follows immediately that all possible versions of this experiment necessarily involve superpositions of quantities which vary under some symmetry transformation, and this entails that the reference system loophole will work for any possible realization of this experiment.

Note that it is important here that although the quantity Alice measures is a partial observable, the \emph{result} of the measurement is a complete observable and so it always has a well-defined value. One consequence of this is that although I have presumed above that Bob can  find out the result of Alice's measurement by some kind of interaction with her (e.g. by means of a conversation), this cannot simply be modelled as a direct measurement on the variable representing the result of the measurement, since complete observables cannot be directly measured. Rather we must suppose that Bob measures some partial observable in an arrangement such that the complete observable associated with the result of his measurement will be appropriately correlated with the complete observable associated with the result of Alice's measurement. More work is needed to establish the details of such a procedure - however, this is not a problem for the present proposal specifically, for if it is true that we only measure partial observables, then one way or another the theory of observability in GR is going to have to reckon with the question of how  observers manage to communicate measurement results corresponding to complete observables to one another. Thus I defer this issue as an interesting question for future work on the partial/complete observables formalism.

Now, at this juncture one might object that the correct resolution to the Wigner's Friend paradoxes should not depend on appeals to relativistic physics or putative future physics. But I am not very moved by this objection, because I see no reason to think that all the puzzles of quantum mechanics can be resolved in a way which is entirely internal to non-relativistic quantum mechanics. Different parts of physics are always in interaction with one another, and since quantum mechanics takes place in a spacetime arena the physics of spacetime is clearly relevant to it. In particular, we should be cognizant of the possibility that spacetime physics might very well turn out to be relevant to quantum mechanics in ways that are not immediately obvious based on ordinary applications of non-relativistic quantum mechanics. 

Indeed, I would argue that appealing to ideas from relativistic physics and quantum gravity is actually a virtue of this approach. For one of the main reasons why one might think it is important to  resolve the puzzles of quantum mechanics is precisely because we have faced challenges in unifying quantum mechanics with relativistic physics, and it is plausible that  these difficulties arise in part because there are  aspects of the standard non-relativistic quantum formalism that we do not yet fully understand, as instanced by the measurement problem. Thus it makes sense to try to resolve these puzzles in ways which bring together quantum physics and relativistic physics, because such  approaches seem more likely to be fruitful with regard to ongoing questions about unification.

\subsection{More complex examples}

One might also try to block the argument from  diffeomorphism invariance by proposing a version of the experiment using variables which have some kind of timeless non-relative significance. For example, we could imagine doing the experiment with the system $S_A$ in a  superposition of messages saying `good' and `bad' - for  of course when Alice reads `good' or `bad' she knows what these words mean without comparing them to a reference system.

However, of course to actually perform such an experiment the words `good' and `bad' must be represented by some kind of physical encoding; thus in the end the association with semantic concepts is irrelevant and what matters is simply whether the encoding variables themselves are variant or invariant. For example, we could imagine just using a spin system and associating `good' with `spin up' and `bad' with `spin down' and then we recover the original experiment as described in section \ref{loophole}: it makes no difference to any part of this story that Alice and Bob have agreed to associate the spin values with semantic concepts. Or we could have some kind of printing machine which goes into a superposition of printing `good' and printing `bad' - but then again, the end result is just two different configurations of points of ink on a page where they stand in different relative distances from each other, so this simply ends up as a very complicated version of the relative distance case as described above, and the same considerations apply. Either way, the association with the concepts `good' and `bad' does not change anything about the experiment.

Finally, one might imagine  bringing in a third observer to be the system $S_A$ and performing the experiment with Alice measuring a superposition of different conscious states of the observer. It seems natural to say that the variables on which different conscious states supervene do not need to be relativized to an external reference system, so a version of the Bong et al experiment performed with such a superposition could not be resolved by the reference system loophole. However, of course nobody has ever actually demonstrated a superposition of distinct conscious states, so the proponent of the reference system loophole can just deny that such superpositions are possible. Additionally, it is natural to suppose that in the end conscious experiences supervene on complete observables, so if it is true that we can only measure partial observables, then it would not be possible for Alice to actually measure the variables on which the  conscious experiences of the relevant observer supervene; rather Alice would always end up measuring some kind of correlated partial observable, in which case the reference frame loophole could still be applied \ref{symmetry}.

 \section{Why this approach? \label{why}}

 We have seen that leveraging the reference system loophole provides a way to resolve the Bong et al  paradox which allows us to maintain the correctness of the quantum predictions (with respect to the appropriate set of quantum degrees of freedom) without sacrificing Local Friendliness. In this section I will turn to the question of why one might consider adopting this resolution to the EWF problem rather than another. 

 As already discussed in section \ref{Wigner}, there are very strong reasons why we should be cautious about denying absoluteness of observed events \citep{https://doi.org/10.48550/arxiv.2203.16278,adlamnonabsolute} - no one has yet demonstrated that this idea can even be stated in a coherent way, let alone showing how to make its epistemology work. Given the very substantive conceptual worries, denying absoluteness of observed events should certainly be a last resort and thus the approach set out here, if it could be made to work, would clearly be preferable.

Additionally, it is possible that some people find the denial of  absoluteness of observed events appealing because they have the intuition that in the EWF scenario there is some kind of indefiniteness of Alice's outcome from Bob's point of view. And  the reference system loophole provides a way of realising this intuition  without jeopardizing baseline scientific presuppositions about memory, experience and communication.  For example, suppose that at the time of the measurement, Alice  says to herself the words `I am currently seeing the result `up' relative to my axis.'  Then the reference system loophole affirms that at later times she will say to herself  `I remember seeing the result `up' relative to my axis' and in the case where Bob asks Alice about her outcome he will hear Alice say the words `I saw the result `up' relative to my axis,' and all of these words mean the same thing to everyone throughout the whole experiment, in an absolute sense; memory and communication  work in exactly the way we would ordinarily expect.  But at the same time,  only when Alice interacts with Bob in the right way does it become fixed what the direction `up relative to Alice's axis' corresponds to in \emph{Bob's} frame of reference, so it is true in quite a strong sense that from Bob's point of view her outcome is really indefinite. Thus the reference system loophole is potentially an appealing intermediate option for those who have considered denying absoluteness of observed events because of intuitions about indefiniteness.

Meanwhile, it is perhaps less obvious that there is anything really wrong with getting rid of the locality assumption or the no-superdeterminism assumption. In particular, one way to deny no-superdeterminism is simply to allow retrocausality, and there are already independent arguments for allowing retrocausality in the context of quantum mechanics \citep{2010arXiv1002.0906P,RevModPhys.92.021002,adlam2022roads} so that seems like quite a natural route to explore. Thus I do not necessarily want to argue that the reference system loophole is preferable to those kinds of options; but nonetheless,  significant numbers of people  are sufficiently determined to maintain locality and no-superdeterminism that they have seriously contemplated denying absoluteness of observed events as an alternative, so the reference system loophole is perhaps most likely to  appeal to people in this group.

Finally, one might also wonder why we should prefer utilizing the reference system loophole over simply adopting an alternative solution to the measurement problem, such as a spontaneous collapse model which avoids the EWF paradox by  biting the bullet and accepting that the correlations across the pairs $\{ (B,D), (B, C),$ $ (A, D), (A, C) \}$ as observed by Bob at the experiment simply will not all obey the predictions of quantum mechanics, even interpreted in a first-person way. The primary reason to prefer the reference system loophole over this is simply that in many quarters it is seen as desirable to maintain the usual quantum predictions as far as possible, and the reference system loophole allows us to retain the quantum predictions to a much greater degree than the spontaneous collapse approach. In particular, the reference system loophole ensures  that in the EWF experiment,  some of Alice's degrees of freedom   remain in superposition relative to Bob, and Bob is able to probe this superposition experimentally and see results corresponding to the usual quantum predictions. By contrast, in a spontaneous collapse approach none of Alice's degrees of freedom remain superposed relative to Bob after her measurement, so he will never see the usual quantum entanglement correlations. Thus the reference system loophole allows us to maintain unitary evolution \emph{in the relevant quantum variables relative to Bob}, which is presumably appealing to anyone who thinks it is generally a good thing to retain the quantum predictions as much as possible.

\subsection{Other EWF Scenarios \label{EWF}}

In this article I have largely focused on the Bong et al EWF experiment, but  it is natural to ask whether a similar approach can be used for other EWF experiments. In fact it can, and indeed such a loophole is actually even more straightforward to implement for the other EWF experiments.

It is helpful to classify EWF scenarios into `first-person' paradoxes and `third-person' paradoxes \citep{adlam2025kindrelationalitydoesquantum}. The first-person paradoxes involve a theorem showing that some set of assumptions together entail  a violation of the predictions of quantum mechanics \emph{within the observations made by a single person} - for example in the Bong et al case, the claim is that Local Friendliness predicts a violation of the quantum predictions as observed just by Bob alone. Consequently a  paradox like this cannot be defused simply by the insistence that the predictions of quantum mechanics  apply only  in a first-person sense.  The Bong et al paradox is the main example of a first-person paradox in the literature, which is why I have focused on it here. 

By contrast, in the third-person paradoxes we instead have a theorem showing that some set of assumptions together entail a violation of the predictions of quantum mechanics across a pair of measurements such that these measurements are performed by different people and  cannot subsequently be compared. For example, the theorem of \cite{ormrod2022nogo} also considers two experiments performed by Alice and Chidi inside  closed laboratories and then supermeasurements made by Bob and Divya, and it uses a similar argument to show that, if we assume absoluteness of observed events, then the correlations between at least one of the pairings of Alice and Chidi, Alice and Divya, Bob and Chidi and Bob and Divya must fail to conform to the predictions of quantum mechanics. But since the measurements made by Bob and Divya destroy the records of the outcomes obtained by Alice and Chidi, it will not be possible for anyone to actually compare these pairs to see the correlations that violate the quantum predictions, so this is only a third-person paradox and not a first-person paradox. As pointed out by \cite{schmid2023review} and \cite{adlam2025kindrelationalitydoesquantum},  the third-person paradoxes can straightforwardly be defused by insisting that the predictions of quantum mechanics apply only in a first-person sense.  

 In this context, the importance of the reference system loophole is that it can provide  a principled reason why the  correlations between variables that are not compared to each other might fail to satisfy the predictions of quantum mechanics. The reference system loophole suggests that this  arises precisely  because, for the same reasons as in the Bong et al case, these measurements do not all take place in the same frame and thus the standard predictions of quantum mechanics for measurements all performed in the same frame do not apply to them. That is, quantum mechanics is not really wrong here, but rather it is simply not applicable to this particular set of outcomes. 

Moreover, the reference system loophole also helps us overcome a natural objection. In the scenario just described it is always possible that Bob and Divya fail to make their supermeasurements and instead open the lab and look inside. In that case the outcomes obtained by Alice and Chidi \emph{can} be compared, and of course we anticipate that when compared they will conform to the predictions of quantum mechanics. So one might  argue that if we make a no-retrocausality assumption such that the results obtained by Alice and Chidi can't depend on whether Bob and Divya actually make their supermeasurements, it follows that these results must still conform to the predictions of quantum mechanics even in the case where the supermeasurements are made. That is, ruling out retrocausality provides a natural reason to suppose that pairs of  measurements should exhibit the usual quantum correlations even when they can never be directly compared.

Now, of course one way to avoid this conclusion is to simply allow retrocausality - but another option is the reference system loophole. That is, suppose we say that as in the Bong et al case, the  reference systems of Alice and Chidi end up in an indefinite state relative to the external laboratories of Bob and Divya after their measurements. Then, just as in the Bong et al case, the correlations between Alice and Chidi, Alice and Divya, Bob and Chidi and Bob and Divya do not have to conform to the predictions of quantum mechanics, because if the supermeasurement is not performed on Alice  then the correlations will be `fixed' retrospectively by the measurement which determines the orientation of Alice's reference system relative to Bob's, and likewise for Chidi. Thus the reference system loophole provides a satisfying way of realizing the intuition that correlations between measurements which can never be directly compared do not have to conform to the predictions of quantum mechanics.

\section{Conclusion}

We have seen here that distinguishing between the observed result of Alice's experiment versus the correct statement of her outcome relative to Bob's external reference system can at least formally allow us to explain the EWF experiment without losing absoluteness of observed events, locality, no-superdeterminism or violating the quantum predictions for the four outcome pairs  $\{ (B,D), (B, C), $ $(A, D), (A, C) \}$. We have also seen that this approach makes sense given our existing understanding of the role of symmetries and invariant properties in such experiments, and that it ties in with ongoing work in relativistic physics and quantum gravity.  

Additionally, this way of thinking about Wigner's Friend and quantum superpositions suggests some interesting new routes for future work on the measurement problem, because it points to an  approach in which only `partial observables' or quantities which vary under a symmetry transformation are quantum degrees of freedom. This is relevant to the measurement problem because it is natural to think that conscious experiences must supervene on invariants or complete observables (since we commonly say that symmetry-variant properties are not physically real by themselves, and within a physicalist view consciousness must supervene on something that is physically real). So if this approach could be made to work, we would never have to worry about superpositions of different conscious states, and thus we would end up with a natural explanation for why observers always have single definite experiences. Particularly if one takes the view that the existing mainstream solutions to the measurement problem are unlikely to be successful \citep{Adlam2025-ADLSSF}, this seems like an interesting new direction to investigate. 

 \section{Acknowledgements}

 Thanks to Anne-Catherine de la Hamette and Andrea di Biagio for discussions about this paper. This work was supported by the  John Templeton Foundation Grant ID 63209.

\end{document}